\documentclass[12pt]{article}

\pdfoutput=1
\usepackage{ifpdf}
\usepackage[a4paper,top=3cm,width=17.4cm, bottom=3.0cm, left=1.8cm]{geometry}
\usepackage{graphicx}
\usepackage[utf8x]{inputenc}

\usepackage{microtype}
\usepackage{ae}
\usepackage{aecompl}

\usepackage{amsmath}
\usepackage{amssymb}
\usepackage{mathdots}

\usepackage{setspace}
\usepackage[small,sf,bf]{caption}
\usepackage[english]{babel}
\usepackage{datetime}
\shortdate 
\usepackage{bm}
\usepackage{subfigure}
\usepackage[colorlinks=true,linktocpage=true,linkcolor=blue,citecolor=blue]{hyperref}
\usepackage[usenames,dvipsnames]{color}

\usepackage[normalem]{ulem} 
\usepackage{soul} 
\usepackage{nicefrac}
\usepackage{float}
\usepackage[titletoc,title]{appendix}
\usepackage{authblk}
\usepackage{cite}

\newcommand{\bel}[1]{\begin{eqnarray}\label{#1}}
\newcommand{\be}{\begin{eqnarray}}
\newcommand{\ee}{\end{eqnarray}}

\newcommand{\rf}[1]{Eq.~(\ref{#1})}
\newcommand{\rfn}[1]{~(\ref{#1})}

\newcommand{\rff}[1]{Fig.~\ref{#1}}

\newcommand{\ackno}{\section*{Acknowledgements}}

\newcommand{\f}[2]{\frac{#1}{#2}}

\newcommand{\sym}{${\mathcal N}=4$}
\newcommand{\symm}{${\mathcal N}=4$ SYM}
\newcommand{\hydron}{hydrodynamization}

\newcommand{\ed}{{\cal E}}       
\newcommand{\peq}{{\cal P}}     
\newcommand{\pT}{{\cal P}_T}   
\newcommand{\pL}{{\cal P}_L}   

\newcommand{\tpi}{\tau_{\pi}}

\newcommand{\pa}{{\cal A}}
\newcommand{\pac}{{a}}

\begin{document}

\title{{\bf Universal behaviour, transients and attractors\\ in supersymmetric
    Yang-Mills plasma}}
\renewcommand\Authfont{\scshape\small}
\renewcommand\Affilfont{\itshape\small}

\author[1,2]{Micha\l{} Spali\'nski}

\affil[1]{Physics Department, University of Bia\l{}ystok, PL-15-245 Bia\l{}ystok, Poland}
\affil[2]{National Center for Nuclear Research, PL-00-681 Warsaw, Poland}

\date{}
\maketitle
\thispagestyle{empty}

\begin{abstract}

Numerical simulations of expanding plasma based on the AdS/CFT correspondence as
well as kinetic theory and hydrodynamic models strongly suggest that some
observables exhibit universal behaviour even when the system is not close to
local equilibrium. This leading behaviour is expected to be corrected by transient,
exponentially decaying contributions which carry information about the initial
state. Focusing on late times, when the system is already
in the hydrodynamic regime, we analyse numerical solutions describing expanding plasma of
strongly coupled \sym\ supersymmetric Yang-Mills theory and identify these
transient effects, matching them in a quantitative way to leading transseries
corrections corresponding to least-damped quasinormal modes of AdS black branes.
In the process we offer additional evidence supporting the recent identification
of the Borel sum of the hydrodynamic gradient expansion with the
far-from-equilibrium attractor in this system.

\end{abstract}


\section{Introduction}

Given the critical role of hydrodynamics in the physical picture behind the
evolution of quark-gluon plasma created in heavy ion collisions, it is very
important to have a reliable understanding of how relativistic systems in highly
nonequilibrium initial states tend toward local equilibrium. It is useful to
view this process as consisting of two stages. The first is characterised by the
quasi-exponential decay of transient effective degrees of freedom -- the
nonhydrodynamic modes, which in the case of \sym\ supersymmetric Yang-Mills
theory (SYM) plasma are in one-to-one correspondence with the quasinormal modes
of AdS black  branes~\cite{Horowitz:1999jd,Kovtun:2005ev}. The second stage is
dominated by slowly-evolving hydrodynamic modes, whose time evolution is
determined by conservation laws. This pattern holds both in microscopic theories
and at the level of hydrodynamic models, such as the M\"uller-Israel-Stewart
(MIS) theory~\cite{Muller:1967zza,Israel:1976tn} and its variants.

This qualitative picture can be made quantitative and appreciated most clearly
when examining certain special observables which enjoy universal behaviour at
late times: they evolve (up to exponentially small corrections) in a way
determined solely by some microscopic model and independently of the initial
conditions. Such quantities, to which one may refer as {\em universal
observables}, exhibit distinct attractor behaviour: for a wide range of initial
conditions numerical solutions tend to a distinguished attractor solution in an
approximately exponential fashion~\cite{Heller:2015dha}. This universal
behaviour often sets in while the system is still highly non-isotropic, and at
later times coincides with the prediction of Navier-Stokes hydrodynamics. It is
natural to interpret \hydron\ in the sense of
Refs.~\cite{Heller:2011ju,Jankowski:2014lna} as a manifestation of this
phenomenon.

This set of ideas leads to the notion that the decay of transient modes describes the way the system approaches a
far-from-equilibrium attractor~\cite{Heller:2015dha,Romatschke:2016hle,Romatschke:2017vte,Florkowski:2017olj} which can
be thought of as ``hydrodynamics beyond the gradient expansion'' in the sense of Refs.~\cite{Lublinsky:2007mm}
and~\cite{Heller:2015dha}. The simplest and most widely studied example appears in the context of conformal Bjorken
flow, where the pressure anisotropy turns out to have such universal behaviour when expressed as a function of the
dimensionless variable $w\equiv\tau T$ (which is the proper time in units of the shear-stress relaxation time, up to a
constant factor). This behaviour is captured by the gradient expansion of the pressure anisotropy, which in this
particular case takes the form of a series in $1/w$. More general expressions for such universal observables -- valid
also without assuming conformal symmetry or boost invariance --  were recently described by
Romatschke~\cite{Romatschke:2017acs}.

An important point is that the gradient expansion itself does not depend on the initial state of the system. However,
this asymptotic solution  receives exponentially-suppressed corrections which on the one hand reflect the spectrum of
nonhydrodynamic modes, and on the other carry information about initial conditions. Specifically, at the linearized
level each black-brane quasinormal mode introduces a transseries sector, which is an infinite series of exponentially
damped corrections, with the damping rate determined by the quasinormal mode frequency. Each such sector enters with an amplitude dependent on the initial state. The entire solution takes the
form of a transseries~\cite{Heller:2015dha,Aniceto:2015mto} whose leading element is the hydrodynamic gradient
expansion.

As first shown in Ref.~\cite{Heller:2013fn}, the hydrodynamic gradient expansion in \symm\ plasma is divergent, and the
precise form of the large order behaviour of the expansion coefficients contains information about the nonhydrodynamic
modes of the system. In fact, each transseries sector contains an asymptotic series in $1/w$. The transseries structure,
together with resurgence relations connecting expansion coefficients in different sectors, gives a consistent solution
when all the divergent power series are properly summed (e.g. using Borel techniques). The Borel summation of the
hydrodynamic gradient series itself gives an approximation of the far-from-equilibrium
attractor~\cite{Spalinski:2017mel}, while the non-trivial transseries sectors describe the dissipation of initial state
information as the attractor is approached. Indeed, the transseries structure -- which was developed in the study of asymptotic series (see  e.g.~\cite{Aniceto:2018bis} and references therein) and has a wide array of applications -- is perfectly suited to capture this phenomenon.

It was shown in Ref.~\cite{Aniceto:2015mto} that the complex-conjugate pair of least-damped quasinormal modes implies a
specific form of the leading transseries correction to the hydrodynamic solution. Thus, in the case of \symm\ the
dominant terms are known precisely, apart from two amplitudes which reflect particular initial conditions. The
calculations reported here are aimed at comparing the form of this leading correction with numerical solutions of the
full time evolution starting from some randomly chosen initial states. Such calculations, based on the AdS/CFT correspondence, were performed in Refs.~\cite{Heller:2011ju,Jankowski:2014lna}. It is well-known that in the first approximation, at sufficiently large times the behaviour of these solutions approaches the prediction of the
leading order of the gradient expansion. The issue addressed here is how the time evolution matches expectations based
on a quantitative analysis of late-time asymptotics.

In a typical initial state many of the nonhydrodynamic modes will be excited, leading to the complex patterns seen at
early times in the solutions of Refs.~\cite{Heller:2011ju,Jankowski:2014lna}. As transients decay, the system tracks the
attractor more and more closely. Soon beyond the regime where \hydron\ occurs, one should expect that the least-damped
nonhydrodynamic modes will become the dominant deviation from purely hydrodynamic behaviour~\cite{Janik:2006gp}. These
subtle traces of the initial state must be present even deep in the hydrodynamic regime. The aim of the work reported
here was to verify that these effects are indeed present. We show how one can isolate these contributions and confirm
that they have the expected form, so one can say that late-time behaviour of the pressure anisotropy describes damped
oscillations around the hydrodynamic attractor. We demonstrate that to see this effect one really needs more than just the gradient expansion truncated at some low order. Finally we compare the attractor of \symm\ to the
attractor of conformal BRSSS hydrodynamics~\cite{Baier:2007ix} and find that the two are essentially identical once the transport coefficients are matched, which may be a part of the reason why relativistic second order hydrodynamics is so successful in the description of quark-gluon
plasma~\cite{Romatschke:2017ejr}.

\section{Universal observables}

The notion of a universal observable is implicit in much of the recent work on
hydrodynamic
attractors~\cite{Heller:2015dha,Romatschke:2016hle,Heller:2016rtz,Romatschke:2017vte,Florkowski:2017olj,Strickland:2017kux,Behtash:2017wqg,Romatschke:2017ejr,Romatschke:2017acs,Almaalol:2018ynx,Denicol:2018pak,Rougemont:2018ivt}, and is rooted in  observations made in Ref.~\cite{Heller:2011ju}.
It is best illustrated in the case of Bjorken flow in conformal BRSSS
hydrodynamics~\cite{Baier:2007ix}. In terms of the longitudinal and transverse
pressures $\pL,\pT$ and equilibrium pressure at the same energy density
$\peq=\ed/3$, the pressure anisotropy is defined by
\bel{rdef}
\pa \equiv \f{\pT-\pL}{\peq} .
\ee
This quantity exhibits universal
behaviour~\cite{Heller:2013fn,Jankowski:2014lna} at late times when expressed
in terms of the dimensionless ``clock variable''
\bel{wdef}
w = \tau T(\tau) .
\ee
This can be inferred directly from
the equations of BRSSS hydrodynamics, which imply the following
equation for the pressure anisotropy $\pa(w)$~\cite{Heller:2015dha,Florkowski:2017olj}:
\bel{feqn}
C_{\tpi}\left(1 + \f{\pa}{12}\right) \pa' + \left(\f{ C_{\lambda_{1}}}{8 C_{\eta}} + \f{C_{\tpi}}{3 w}
\right) \pa^2 = \f{3}{2} \left(\f{8 C_\eta}{w} - \pa\right)
\ee
where $C_\eta\equiv\eta/s$ is the ratio of the shear viscosity to entropy
density, and $C_{\tpi},C_{\lambda_{1}}$ are dimensionless second-order
transport coefficients.  At late times, which corresponds to large values of
$w$, this equation possesses an asymptotic solution of the form
\bel{gradex}
\pa(w) = \sum_{n=1}^{\infty} \pac_n \, w^{-n} =  \f{8 C_\eta}{w} + \dots
\ee
This solution contains no trace of initial conditions, and thus it is
universal in the sense that all solutions tend to it regardless of the initial
state.  One can go beyond the asymptotic solution \rfn{gradex} by including
exponential corrections -- this is the subject of the following section.

The leading term describing the way in which the pressure anisotropy
approaches zero at late times depends on the shear viscosity to entropy ratio,
so it will depend on the physical system under consideration.  However, if one
rescales
the clock variable by introducing~\cite{Heller:2016rtz}
\bel{wtilde} \tilde{w}
\equiv \f{w}{4\pi C_\eta}
\ee
then the leading asymptotic behaviour of the pressure
anisotropy
\bel{palead}
\pa(\tilde{w}) = \f{2}{\pi\tilde{w}} + \dots
\ee
is universal not only in the sense of being independent of initial conditions,
but is also independent of any transport coefficients. The choice of scaling
in \rf{wtilde} is such that the leading term in \rf{palead} takes the form
of \rf{gradex} when the shear viscosity to entropy density ratio
assumes the value $C_\eta=1/4\pi$.

Although the discussion in this section was framed in the context of BRSSS fluid dynamics, it is in fact far more
general. The asymptotic behaviour of universal observables expresses the fact that the predictions of any sensible
theory of hydrodynamics approach those of Navier-Stokes theory at late times. While this is obvious by construction, it
is only {\em apparent} if one examines a universal observable such as $\pa(w)$. In the evolution of non-universal
quantities such as the temperature, this basic fact will be obscured by the dependence on initial conditions.

\section{The form of transients}

The universal behaviour of $\pa(w)$ makes it possible to study transient
effects -- the decay of nonhydrodynamic modes -- in an unambiguous way. Since
the asymptotic expansion given in \rf{gradex} is independent of initial conditions, all
solutions are guaranteed to behave in accordance with it up to exponential
corrections. These effects, at least in hydrodynamic theories and in strongly
coupled \symm\ plasma, are captured by the transseries
representation~\cite{Heller:2015dha,Aniceto:2015mto}. The need to include
transseries corrections is also a reflection of the fact that the gradient
expansion in \rf{gradex} is divergent both at the microscopic
level~\cite{Heller:2013fn,Heller:2016rtz} (see also
Ref.~\cite{Denicol:2016bjh}) and in
hydrodynamics~\cite{Heller:2015dha,Aniceto:2015mto}.

The pressure anisotropy can be written as a sum of two contributions
\bel{pacorr}
\pa =\pa_H + \delta\pa
\ee
where $\pa_H$ represents the purely hydrodynamic, universal part of the
solution and $\delta\pa$ is a correction which will depend on the initial
conditions, and will thus be different for different solutions. Due to the
divergence of the gradient expansion $\pa_H$ has to be understood either in
the sense of a truncation of the gradient series at some low order, or as the
result of a applying a procedure such as Borel summation.

The precise form of the correction depends on the spectrum of nonhydrodynamic modes. Roughly speaking, at the linearized
level each nonhydrodynamic mode makes a contribution of the form
\bel{deltapr}
\delta\pa(w) = \sigma w^\beta e^{-Aw} \Phi(w)
\ee
where $\Phi$ is an infinite series in powers of $1/w$ whose coefficients are determined in terms of the parameters of the theory, as are the constants $A$ and $\beta$. The amplitude $\sigma$ however depends on the initial conditions. In general
all these parameters are complex.

In the case of \symm\ each AdS black-brane quasinormal mode makes a contribution of the
form \rfn{deltapr}, where the parameter $A$ is proportional to the complex QNM
frequency. The precise form of the correction due to the complex-conjugate pair of
least-damped nonhydrodynamic modes can be inferred from
Refs.~\cite{Janik:2006gp,Heller:2013fn,Heller:2014wfa,Aniceto:2015mto} and
reads
\bel{transcor}
\delta\pa(w) \sim e^{-\f{3}{2}\Omega_I w} w^{\beta_R}\left[\Phi_+(w)\cos\left(\f{3}{2}\Omega_R w -\beta_I \log(w)\right)
+ \Phi_-(w) \sin\left(\f{3}{2}\Omega_R w -\beta_I \log(w)\right)\right] .
\ee
In this equation, $\Phi_\pm(w)$ denote infinite series of the
form similar to \rf{gradex} appearing in the first transseries sectors, but
with a constant leading term:
\bel{phis}
\Phi_\pm(w) = \sigma_\pm \left(1 +  \sum_{n=1}^{\infty} a^{(\pm)}_n \, w^{-n}\right) .
\ee
The quantities $\sigma_\pm$ play the role of integration constants.
The remaining coefficients $a^{(\pm)}_n$
are independent of initial conditions, but their specific values
are not relevant for the present study and we will approximate the entire
series in \rf{phis} by constant amplitudes, i.e. by the leading contributions
given by the integration constants $\sigma_\pm$. In what follows we will fit
these amplitudes to numerical data from AdS/CFT simulations.

The values of the remaining parameters appearing in \rf{transcor} are
known. The least-damped AdS black brane QNM frequencies~\cite{Nunez:2003eq}
(corresponding to an operator of conformal weight $\Delta=4$, apart from a
factor of $\pi$) are given by:
\bel{omegas}
\Omega_{R} \approx 9.800 , \qquad  \Omega_{I} \approx 8.629
\ee
and, furthermore
\bel{betas}
\beta_{R} \approx 0.6866 , \qquad  \beta_{I} \approx 0.7798
\ee
which can be extracted from Ref.~\cite{Heller:2013fn}.

There are two types of correction to \rf{transcor}: nonlinear terms, which are
damped by powers of the exponential already appearing there and further
contributions of a similar form, but with parameter values corresponding to
more strongly-damped QNM of the AdS black brane. These contributions are
subleading and will be ignored in the following.

\section{Matching numerical evolution to the leading QNM}

We now turn to examining numerical solutions of Bjorken flow with the
aim of confronting their late time behaviour with the expectations discussed
in the previous section. Suitable solutions can be obtained using the approach
of Ref.~\cite{Jankowski:2014lna} (using methods developed earlier in
Refs.~\cite{Chesler:2009cy,Heller:2011ju}). The focus of that study was on
\hydron, so randomly generated initial conditions were evolved only
until hydrodynamic behaviour was identified -- this occurred overwhelmingly for
$w<1$.  Note that one could contemplate interpreting solutions at times
preceding \hydron\ in terms of compositions of quasinormal modes, in the
spirit of what was done in Refs.~\cite{Heller:2012km,Heller:2013oxa} for the
isotropisation problem.  In the present study however, we are interested in
transient effects {\em after} \hydron, which requires evolving the
asymptotically-AdS geometry to much later times, where only the least damped
pair of quasinormal modes is relevant.  This is based on the observation that
while at early times all the quasinormal modes can a priori play an important
role in a given solution, some time after \hydron\ occurs only the
least-damped QNM should dominate deviations from the asymptotic form given in
\rf{gradex}.

\begin{figure}
\begin{center}
\includegraphics[height=0.36\textheight]{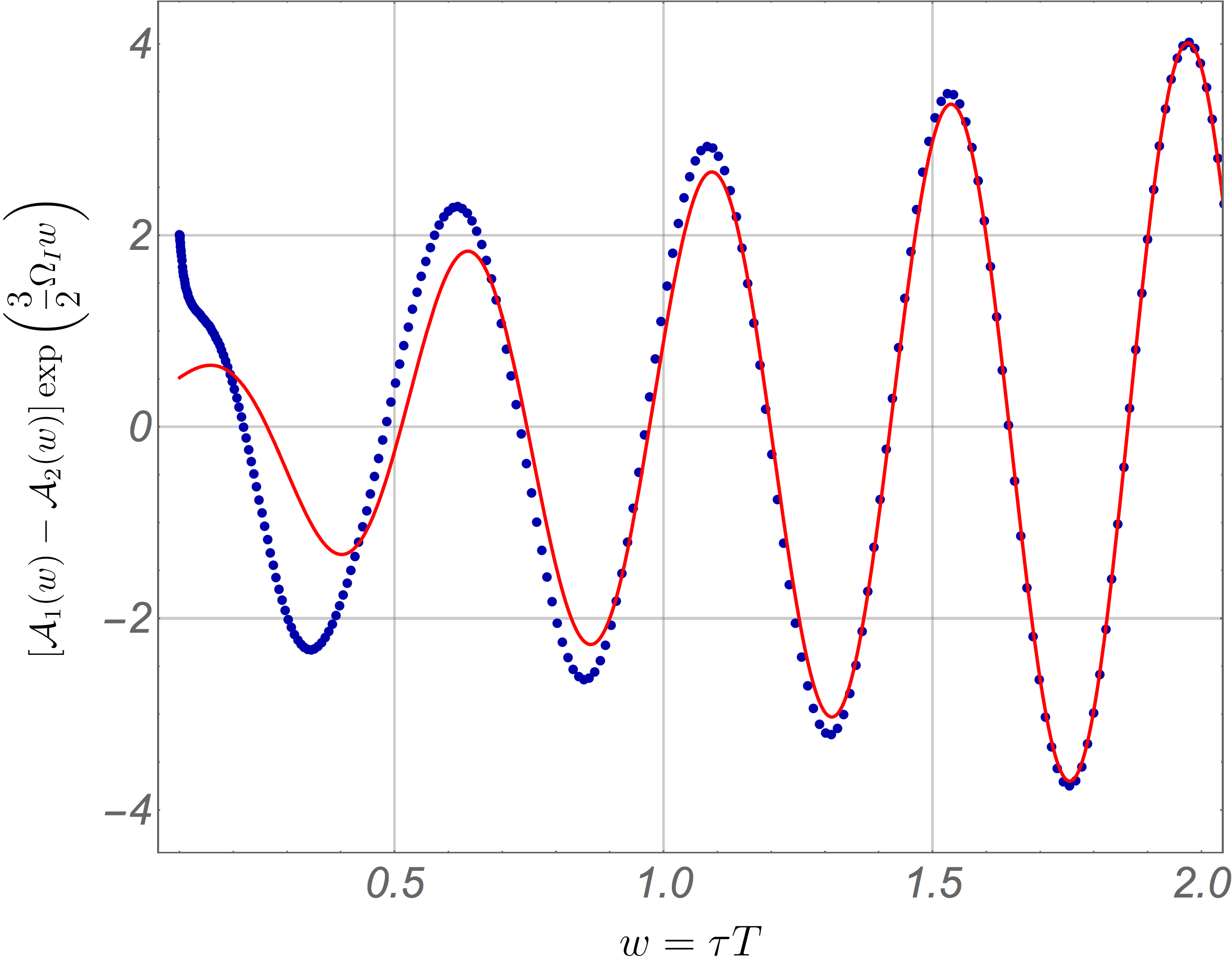}
\caption{The difference between two numerical solutions (blue, dotted) fitted
  to the form in \rf{difference} (red). Only the amplitudes $C, \tilde{C}$ appearing in \rf{difference} are
  fitted. Clearly, for $w<1$ the single QNM correction
  is a very poor approximation, but
  already at $w\approx3/2$ one notes excellent agreement.}
\label{fig:diff}
\end{center}
\end{figure}

To isolate this effect explicitly we will first make use of the observation
also used in Ref.~\cite{Heller:2018qvh}, which is that the universal part
$\pa_H$ will cancel in the difference of any pair of solutions $\pa_1(w)$,
$\pa_2(w)$, so that
\bel{difference}
\pa_1(w)-\pa_2(w)=e^{-\f{3}{2}\Omega_I w} w^{\beta_R}\left[C \cos\left(\f{3}{2}\Omega_R w -\beta_I \log(w)\right) + \tilde{C}  \sin\left(\f{3}{2}\Omega_R w -\beta_I \log(w)\right)\right]
\ee
where the coefficients $C, \tilde{C}$ are given by the differences of amplitudes
$\sigma_\pm$ of the QNM contributions to each of the solutions $\pa_1(w)$,
$\pa_2(w)$. Since apart from these amplitudes all the parameters appearing in
\rf{difference} are known, one can fit the constants $C, \tilde{C}$ and check
if one can reproduce the behaviour of differences of numerical solutions. As
seen from \rff{fig:diff}, at early times this fails, but already at $w>3/2$ it
works very well. This is completely in line with expectations. Only one pair
of solutions was used to generate \rff{fig:diff}, but we considered a number
of such pairs and confirmed that they all lead to the same conclusion.


There is an alternative to the approach described above. Instead of
considering the difference of two numerical solutions one can consider the
difference between a numerical solution and some approximate representation of
the universal hydrodynamic contribution $\pa_H$. As mentioned earlier, there
are two obvious choices here: a truncation of the series in \rf{gradex}, or
its Borel sum. If either of these choices provides an accurate representation
of $\pa_H$ we should find that subtracting it from any numerical solution will
again leave an exponentially-damped correction of the form \rfn{difference}.

If one considers, instead of \rf{difference}, the difference between a
numerical solution and a truncation of the gradient series, one finds a
situation such as that depicted in left plot of \rff{fig:subhydro}, which
results in the case of truncating the gradient expansion at
first-order. Clearly, such an approximation is inadequate for the present
purpose. It turns out that going to second order does not improve the
situation.

The other option is to use the Borel sum of the gradient expansion of \symm\ which was recently computed in
Ref.~\cite{Spalinski:2017mel}. It was also shown there that the Borel sum acts as a far-from-equilibrium attractor for
numerical solutions such as those discussed in the previous section. This fact can be taken as supporting evidence for
the idea that this sum represents ``hydrodynamics beyond the gradient expansion'' and gives motivation for considering
it as a useful estimate of $\pa_H$. Indeed, as seen from the right-hand plot in \rff{fig:subhydro}, subtracting the
Borel sum is, at least qualitatively as effective as subtracting a full numerical solution.

\begin{figure}
\begin{center}
\includegraphics[height=0.27\textheight]{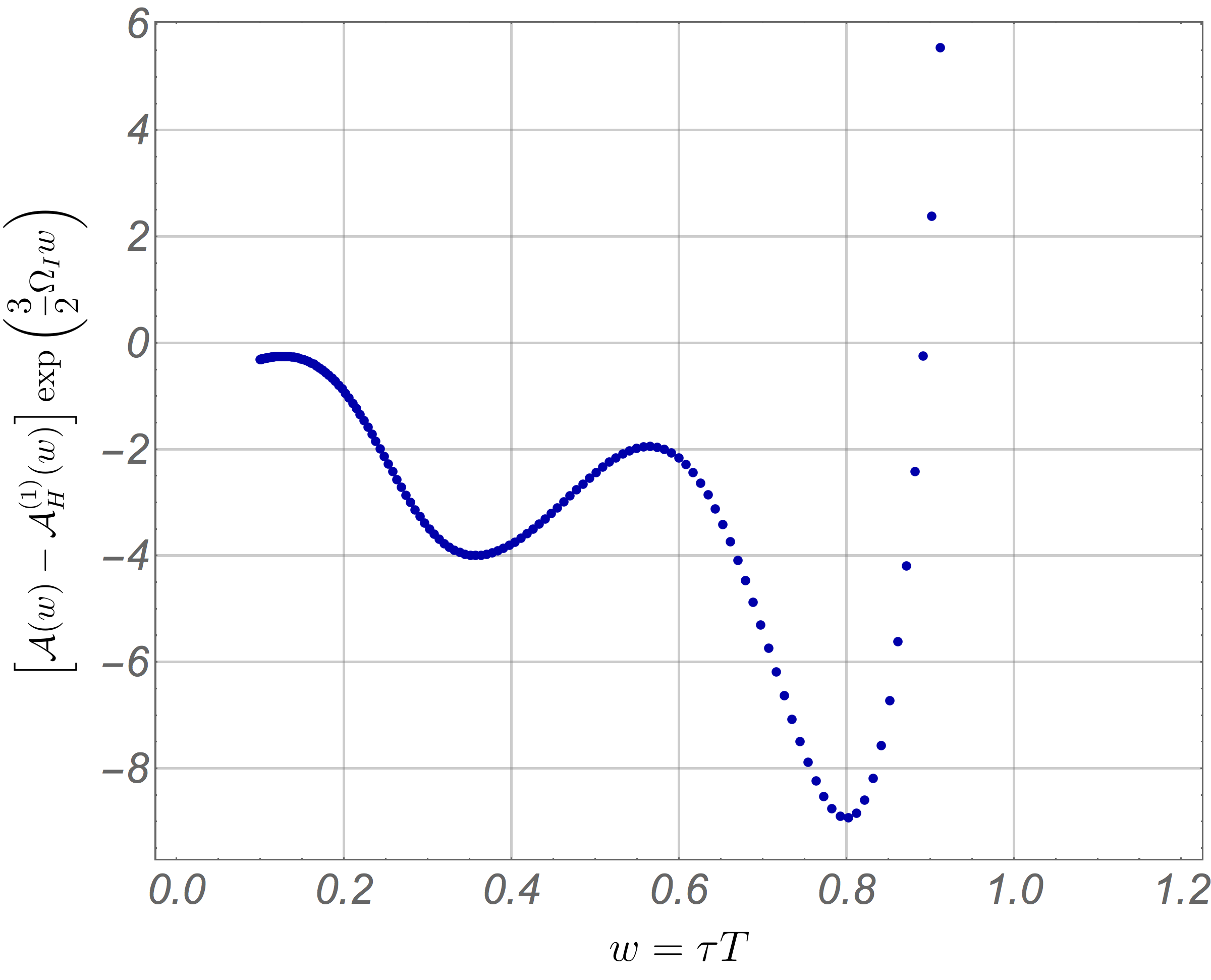}
\includegraphics[height=0.27\textheight]{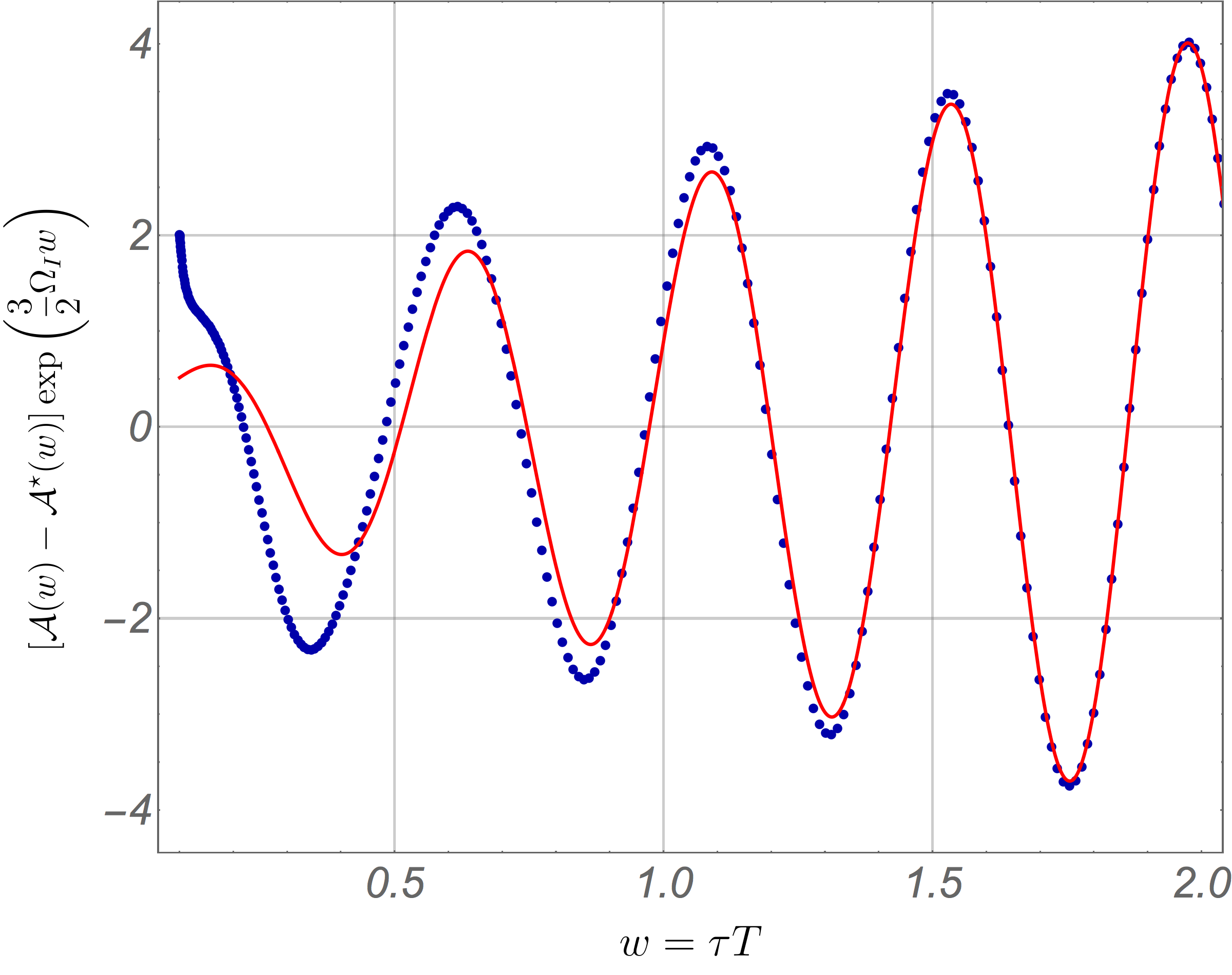}
\caption{
  Left plot: the difference between a numerical solution and the first-order
  truncation of the gradient expansion. Right plot: the difference between a
  numerical solution and the attractor (approximated by the Borel sum). The
  red line is obtained by fitting the amplitudes in \rf{transcor} to the
  numerical data (blue, dotted).}
\label{fig:subhydro}
\end{center}
\end{figure}

Note that if we were to study the difference between some numerical solution and the attractor determined as a solution
of the AdS equations of motion (such as was done in Ref.~\cite{Romatschke:2017vte}) this would just be a special case of
the argument given earlier in this section, since the attractor is (after all) also a solution -- albeit a rather
special one. However, if we instead take the Borel sum of the gradient expansion~\cite{Spalinski:2017mel} this can be
interpreted as a test of whether such a procedure yields a result which is exponentially close to a full numerical
solution.


\section{Conclusions and outlook}

The existence of universal observables which exhibit attractor behaviour and the availability of very precise numerical
simulations of Bjorken flow~\cite{Jankowski:2014lna} makes it possible study transient effects in the late time
behaviour of numerical solutions of the full nonlinear evolution equations based on the AdS/CFT correspondence. The fact
that these effects can be explicitly detected in the numerical simulations supports the general picture of \hydron\
developed in a number of recent
works~\cite{Heller:2013fn,Jankowski:2014lna,Heller:2014wfa,Heller:2015dha,Romatschke:2016hle,Heller:2016rtz,Romatschke:2017vte,Florkowski:2017olj,Strickland:2017kux,Behtash:2017wqg,Romatschke:2017ejr,Almaalol:2018ynx,Denicol:2018pak,Rougemont:2018ivt,Withers:2018srf,Behtash:2018moe}.
It is gratifying that these effects can be found in the precise form expected on the basis of the identification of
transient, nonhydrodynamic modes of the expanding plasma with the quasinormal modes of AdS black branes and the
realization that the hydrodynamic gradient expansion is the leading element of a transseries.

We have also demonstrated that the Borel sum of the gradient series, which was
calculated and shown to act as an attractor in Ref.~\cite{Spalinski:2017mel},
is a useful proxy for the notion of ``hydrodynamics beyond the gradient
expansion''~\cite{Lublinsky:2007mm,Heller:2015dha}. In view of this it is
interesting to ask to what extent the attractor of \symm\ coincides with the
attractor of MIS theory, which is most commonly used to model the hydrodynamic
stage of evolution of quark-gluon plasma created in heavy-ion collisions. The
result of such a comparison is presented in \rff{fig:compare}, where the BRSSS
variant of MIS theory is used, with the shear viscosity, relaxation time and
$\lambda_1$ transport coefficients fitted to their \symm\ values known from
fluid-gravity duality~\cite{Bhattacharyya:2008jc}:
\bel{symvalues}
 C_{\tau \Pi } = \frac{2-\log (2)}{2 \pi} , \qquad  C_{\lambda_1} = \frac{1}{2
     \pi}, \qquad  C_\eta = \frac{1}{4 \pi}.
\ee
\begin{figure}
\begin{center}
\includegraphics[height=0.36\textheight]{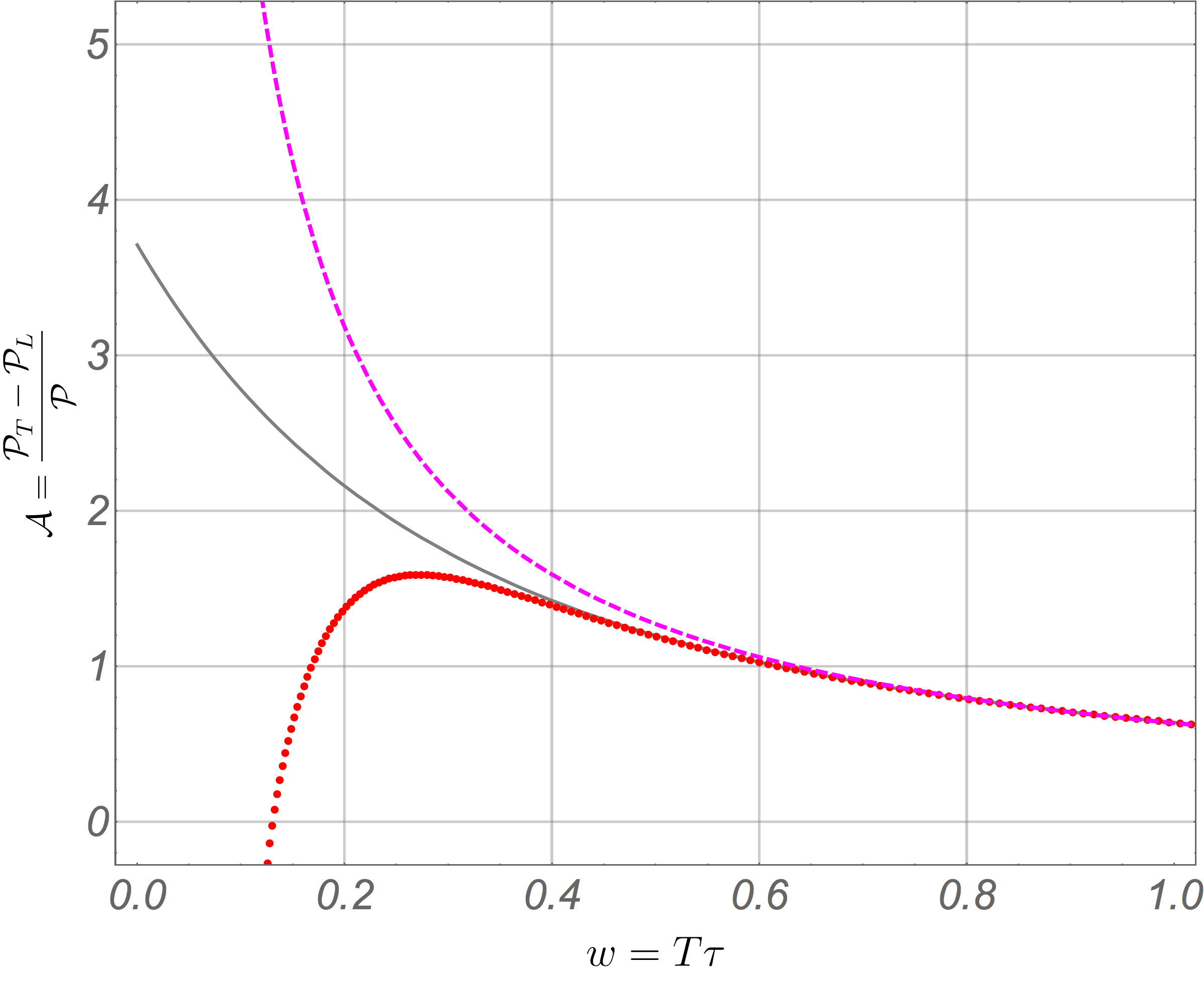}
\caption{The attractor of BRSSS
theory (gray) and the Borel sum of the gradient series of \symm\ (red,
dotted). Also shown in the first order
truncation of the gradient expansion (magenta, dashed).}
\label{fig:compare}
\end{center}
\end{figure}
It is striking that the two attractors coincide almost as soon as the Borel summation can be considered reliable (and
coincides with the result of a calculation reported in Ref.~\cite{Romatschke:2017vte} which was based on a different
approach).  This is in contrast to the attractor of kinetic theory in the relaxation time approximation
(RTA)~\cite{Romatschke:2017vte,Casalderrey-Solana:2017zyh,Blaizot:2017ucy}, which is not very well reproduced by BRSSS
hydrodynamics. Indeed, in a recent study~\cite{Strickland:2017kux} the attractor of RTA kinetic theory was compared to
the MIS attractor, as well as to the attractor of anisotropic hydrodynamics (see
e.g.~Ref.~\cite{Florkowski:2017mgu,Alqahtani:2017jwl}). In that case the MIS attractor was found to converge to the
exact kinetic theory result rather late, with anisotropic hydrodynamics reproducing it well already at very early times.
Similar results were found in Ref.~\cite{Behtash:2017wqg}, where also the DNMR hydrodynamic theory~\cite{Denicol:2012cn}
was considered. In this context note that the recent study of Ref.~\cite{Ghiglieri:2018dgf} suggests that despite
similarities in \hydron~\cite{Heller:2016rtz} the approaches based on kinetic theory and on the AdS/CFT correspondence
are quite far apart by some metrics. It is an open question at the moment which of these two paradigms captures
essential features of quantum chromodynamics more closely.

\ackno
I would like to thank Micha\l\ P. Heller, Jakub Jankowski and Mauricio Martinez for discussions and useful comments on the manuscript. This work was supported by the National Science Centre Grant No. 2015/19/B/ST2/02824.

\bibliographystyle{utphys}
\bibliography{seebib}{}

\providecommand{\href}[2]{#2}\begingroup\raggedright\begin{thebibliography}{10}

\bibitem{Horowitz:1999jd}
G.~T. Horowitz and V.~E. Hubeny, ``{Quasinormal modes of AdS black holes and
  the approach to thermal equilibrium},''
  \href{http://dx.doi.org/10.1103/PhysRevD.62.024027}{{\em Phys. Rev.} {\bf
  D62} (2000)  024027},
\href{http://arxiv.org/abs/hep-th/9909056}{{\tt arXiv:hep-th/9909056
  [hep-th]}}.

\bibitem{Kovtun:2005ev}
P.~K. Kovtun and A.~O. Starinets, ``{Quasinormal modes and holography},''
  \href{http://dx.doi.org/10.1103/PhysRevD.72.086009}{{\em Phys. Rev.} {\bf
  D72} (2005)  086009},
\href{http://arxiv.org/abs/hep-th/0506184}{{\tt arXiv:hep-th/0506184
  [hep-th]}}.

\bibitem{Muller:1967zza}
I.~Muller, ``{Zum Paradoxon der Warmeleitungstheorie},''
\href{http://dx.doi.org/10.1007/BF01326412}{{\em Z. Phys.} {\bf 198} (1967)
  329--344}.

\bibitem{Israel:1976tn}
W.~Israel, ``{Nonstationary irreversible thermodynamics: A Causal relativistic
  theory},''
\href{http://dx.doi.org/10.1016/0003-4916(76)90064-6}{{\em Annals Phys.} {\bf
  100} (1976)  310--331}.

\bibitem{Heller:2015dha}
M.~P. Heller and M.~Spaliński, ``{Hydrodynamics Beyond the Gradient Expansion:
  Resurgence and Resummation},''
  \href{http://dx.doi.org/10.1103/PhysRevLett.115.072501}{{\em Phys. Rev.
  Lett.} {\bf 115} (2015) no.~7, 072501},
\href{http://arxiv.org/abs/1503.07514}{{\tt arXiv:1503.07514 [hep-th]}}.

\bibitem{Heller:2011ju}
M.~P. Heller, R.~A. Janik, and P.~Witaszczyk, ``{The characteristics of
  thermalization of boost-invariant plasma from holography},''
  \href{http://dx.doi.org/10.1103/PhysRevLett.108.201602}{{\em Phys.Rev.Lett.}
  {\bf 108} (2012)  201602},
\href{http://arxiv.org/abs/1103.3452}{{\tt arXiv:1103.3452 [hep-th]}}.

\bibitem{Jankowski:2014lna}
J.~Jankowski, G.~Plewa, and M.~Spaliński, ``{Statistics of thermalization in
  Bjorken Flow},'' \href{http://dx.doi.org/10.1007/JHEP12(2014)105}{{\em JHEP}
  {\bf 12} (2014)  105},
\href{http://arxiv.org/abs/1411.1969}{{\tt arXiv:1411.1969 [hep-th]}}.

\bibitem{Florkowski:2017olj}
W.~Florkowski, M.~P. Heller, and M.~Spaliński, ``{New theories of relativistic
  hydrodynamics in the LHC era},''
  \href{http://dx.doi.org/10.1088/1361-6633/aaa091}{{\em Rept. Prog. Phys.}
  {\bf 81} (2018) no.~4, 046001},
\href{http://arxiv.org/abs/1707.02282}{{\tt arXiv:1707.02282 [hep-ph]}}.

\bibitem{Romatschke:2017acs}
P.~Romatschke, ``{Relativistic Hydrodynamic Attractors with Broken Symmetries:
  Non-Conformal and Non-Homogeneous},''
  \href{http://dx.doi.org/10.1007/JHEP12(2017)079}{{\em JHEP} {\bf 12} (2017)
  079},
\href{http://arxiv.org/abs/1710.03234}{{\tt arXiv:1710.03234 [hep-th]}}.

\bibitem{Aniceto:2015mto}
I.~Aniceto and M.~Spaliński, ``{Resurgence in Extended Hydrodynamics},''
  \href{http://dx.doi.org/10.1103/PhysRevD.93.085008}{{\em Phys. Rev.} {\bf
  D93} (2016) no.~8, 085008},
\href{http://arxiv.org/abs/1511.06358}{{\tt arXiv:1511.06358 [hep-th]}}.

\bibitem{Janik:2006gp}
R.~A. Janik and R.~B. Peschanski, ``{Gauge/gravity duality and thermalization
  of a boost-invariant perfect fluid},''
  \href{http://dx.doi.org/10.1103/PhysRevD.74.046007}{{\em Phys. Rev.} {\bf
  D74} (2006)  046007},
\href{http://arxiv.org/abs/hep-th/0606149}{{\tt arXiv:hep-th/0606149
  [hep-th]}}.

\bibitem{Lublinsky:2007mm}
M.~Lublinsky and E.~Shuryak, ``{How much entropy is produced in strongly
  coupled Quark-Gluon Plasma (sQGP) by dissipative effects?},''
  \href{http://dx.doi.org/10.1103/PhysRevC.76.021901}{{\em Phys. Rev.} {\bf
  C76} (2007)  021901},
\href{http://arxiv.org/abs/0704.1647}{{\tt arXiv:0704.1647 [hep-ph]}}.

\bibitem{Spalinski:2017mel}
M.~Spaliński, ``{On the hydrodynamic attractor of Yang–Mills plasma},''
  \href{http://dx.doi.org/10.1016/j.physletb.2017.11.059}{{\em Phys. Lett.}
  {\bf B776} (2018)  468--472},
\href{http://arxiv.org/abs/1708.01921}{{\tt arXiv:1708.01921 [hep-th]}}.

\bibitem{Romatschke:2016hle}
P.~Romatschke, ``{Do nuclear collisions create a locally equilibrated
  quark-gluon plasma?},''
  \href{http://dx.doi.org/10.1140/epjc/s10052-016-4567-x}{{\em Eur. Phys. J.}
  {\bf C77} (2017) no.~1, 21},
\href{http://arxiv.org/abs/1609.02820}{{\tt arXiv:1609.02820 [nucl-th]}}.

\bibitem{Heller:2016rtz}
M.~P. Heller, A.~Kurkela, M.~Spaliński, and V.~Svensson, ``{Hydrodynamization
  in kinetic theory: transient modes and the gradient expansion},'' {\em Phys.
  Rev. D, to appear} (2016)  ,
\href{http://arxiv.org/abs/1609.04803}{{\tt arXiv:1609.04803 [nucl-th]}}.

\bibitem{Romatschke:2017vte}
P.~Romatschke, ``{Relativistic Fluid Dynamics Far From Local Equilibrium},''
  \href{http://dx.doi.org/10.1103/PhysRevLett.120.012301}{{\em Phys. Rev.
  Lett.} {\bf 120} (2018) no.~1, 012301},
\href{http://arxiv.org/abs/1704.08699}{{\tt arXiv:1704.08699 [hep-th]}}.

\bibitem{Strickland:2017kux}
M.~Strickland, J.~Noronha, and G.~Denicol, ``{Anisotropic nonequilibrium
  hydrodynamic attractor},''
  \href{http://dx.doi.org/10.1103/PhysRevD.97.036020}{{\em Phys. Rev.} {\bf
  D97} (2018) no.~3, 036020},
\href{http://arxiv.org/abs/1709.06644}{{\tt arXiv:1709.06644 [nucl-th]}}.

\bibitem{Behtash:2017wqg}
A.~Behtash, C.~N. Cruz-Camacho, and M.~Martinez, ``{Far-from-equilibrium
  attractors and nonlinear dynamical systems approach to the Gubser flow},''
  \href{http://dx.doi.org/10.1103/PhysRevD.97.044041}{{\em Phys. Rev.} {\bf
  D97} (2018) no.~4, 044041},
\href{http://arxiv.org/abs/1711.01745}{{\tt arXiv:1711.01745 [hep-th]}}.

\bibitem{Romatschke:2017ejr}
P.~Romatschke and U.~Romatschke, ``{Relativistic Fluid Dynamics In and Out of
  Equilibrium -- Ten Years of Progress in Theory and Numerical Simulations of
  Nuclear Collisions},''
\href{http://arxiv.org/abs/1712.05815}{{\tt arXiv:1712.05815 [nucl-th]}}.

\bibitem{Almaalol:2018ynx}
D.~Almaalol and M.~Strickland, ``{Anisotropic hydrodynamics with a scalar
  collisional kernel},''
  \href{http://dx.doi.org/10.1103/PhysRevC.97.044911}{{\em Phys. Rev.} {\bf
  C97} (2018) no.~4, 044911},
\href{http://arxiv.org/abs/1801.10173}{{\tt arXiv:1801.10173 [hep-ph]}}.

\bibitem{Denicol:2018pak}
G.~S. Denicol and J.~Noronha, ``{Hydrodynamic attractor and the fate of
  perturbative expansions in Gubser flow},''
\href{http://arxiv.org/abs/1804.04771}{{\tt arXiv:1804.04771 [nucl-th]}}.

\bibitem{Rougemont:2018ivt}
R.~Rougemont, R.~Critelli, and J.~Noronha, ``{Non-hydrodynamic quasinormal
  modes and equilibration of a baryon dense holographic QGP with a critical
  point},''
\href{http://arxiv.org/abs/1804.00189}{{\tt arXiv:1804.00189 [hep-ph]}}.

\bibitem{Baier:2007ix}
R.~Baier, P.~Romatschke, D.~T. Son, A.~O. Starinets, and M.~A. Stephanov,
  ``{Relativistic viscous hydrodynamics, conformal invariance, and
  holography},'' \href{http://dx.doi.org/10.1088/1126-6708/2008/04/100}{{\em
  JHEP} {\bf 04} (2008)  100},
\href{http://arxiv.org/abs/0712.2451}{{\tt arXiv:0712.2451 [hep-th]}}.

\bibitem{Heller:2013fn}
M.~P. Heller, R.~A. Janik, and P.~Witaszczyk, ``{Hydrodynamic Gradient
  Expansion in Gauge Theory Plasmas},''
  \href{http://dx.doi.org/10.1103/PhysRevLett.110.211602}{{\em Phys.Rev.Lett.}
  {\bf 110} (2013) no.~21, 211602},
\href{http://arxiv.org/abs/1302.0697}{{\tt arXiv:1302.0697 [hep-th]}}.

\bibitem{Denicol:2016bjh}
G.~S. Denicol and J.~Noronha, ``{Divergence of the Chapman-Enskog expansion in
  relativistic kinetic theory},''
\href{http://arxiv.org/abs/1608.07869}{{\tt arXiv:1608.07869 [nucl-th]}}.

\bibitem{Heller:2014wfa}
M.~P. Heller, R.~A. Janik, M.~Spaliński, and P.~Witaszczyk, ``{Coupling
  hydrodynamics to nonequilibrium degrees of freedom in strongly interacting
  quark-gluon plasma},''
  \href{http://dx.doi.org/10.1103/PhysRevLett.113.261601}{{\em Phys.Rev.Lett.}
  {\bf 113} (2014) no.~26, 261601},
\href{http://arxiv.org/abs/1409.5087}{{\tt arXiv:1409.5087 [hep-th]}}.

\bibitem{Nunez:2003eq}
A.~Nunez and A.~O. Starinets, ``{AdS / CFT correspondence, quasinormal modes,
  and thermal correlators in N=4 SYM},''
  \href{http://dx.doi.org/10.1103/PhysRevD.67.124013}{{\em Phys. Rev.} {\bf
  D67} (2003)  124013},
\href{http://arxiv.org/abs/hep-th/0302026}{{\tt arXiv:hep-th/0302026
  [hep-th]}}.

\bibitem{Chesler:2009cy}
P.~M. Chesler and L.~G. Yaffe, ``{Boost invariant flow, black hole formation,
  and far-from-equilibrium dynamics in N = 4 supersymmetric Yang-Mills
  theory},'' \href{http://dx.doi.org/10.1103/PhysRevD.82.026006}{{\em
  Phys.Rev.} {\bf D82} (2010)  026006},
\href{http://arxiv.org/abs/0906.4426}{{\tt arXiv:0906.4426 [hep-th]}}.

\bibitem{Heller:2012km}
M.~P. Heller, D.~Mateos, W.~van~der Schee, and D.~Trancanelli, ``{Strong
  Coupling Isotropization of Non-Abelian Plasmas Simplified},''
  \href{http://dx.doi.org/10.1103/PhysRevLett.108.191601}{{\em Phys. Rev.
  Lett.} {\bf 108} (2012)  191601},
\href{http://arxiv.org/abs/1202.0981}{{\tt arXiv:1202.0981 [hep-th]}}.

\bibitem{Heller:2013oxa}
M.~P. Heller, D.~Mateos, W.~van~der Schee, and M.~Triana, ``{Holographic
  isotropization linearized},''
  \href{http://dx.doi.org/10.1007/JHEP09(2013)026}{{\em JHEP} {\bf 09} (2013)
  026},
\href{http://arxiv.org/abs/1304.5172}{{\tt arXiv:1304.5172 [hep-th]}}.

\bibitem{Heller:2018qvh}
M.~P. Heller and V.~Svensson, ``{How does relativistic kinetic theory remember
  about initial conditions?},''
\href{http://arxiv.org/abs/1802.08225}{{\tt arXiv:1802.08225 [nucl-th]}}.

\bibitem{Withers:2018srf}
B.~Withers, ``{Short-lived modes from hydrodynamic dispersion relations},''
\href{http://arxiv.org/abs/1803.08058}{{\tt arXiv:1803.08058 [hep-th]}}.

\bibitem{Bhattacharyya:2008jc}
S.~Bhattacharyya, V.~E. Hubeny, S.~Minwalla, and M.~Rangamani, ``{Nonlinear
  Fluid Dynamics from Gravity},''
  \href{http://dx.doi.org/10.1088/1126-6708/2008/02/045}{{\em JHEP} {\bf 0802}
  (2008)  045},
\href{http://arxiv.org/abs/0712.2456}{{\tt arXiv:0712.2456 [hep-th]}}.

\bibitem{Casalderrey-Solana:2017zyh}
J.~Casalderrey-Solana, N.~I. Gushterov, and B.~Meiring, ``{Resurgence and
  Hydrodynamic Attractors in Gauss-Bonnet Holography},''
  \href{http://dx.doi.org/10.1007/JHEP04(2018)042}{{\em JHEP} {\bf 04} (2018)
  042},
\href{http://arxiv.org/abs/1712.02772}{{\tt arXiv:1712.02772 [hep-th]}}.

\bibitem{Blaizot:2017ucy}
J.-P. Blaizot and L.~Yan, ``{Fluid dynamics of out of equilibrium boost
  invariant plasmas},''
  \href{http://dx.doi.org/10.1016/j.physletb.2018.02.058}{{\em Phys. Lett.}
  {\bf B780} (2018)  283--286},
\href{http://arxiv.org/abs/1712.03856}{{\tt arXiv:1712.03856 [nucl-th]}}.

\bibitem{Florkowski:2017mgu}
W.~Florkowski, ``{Various Approaches to Anisotropic Hydrodynamics},''
\href{http://dx.doi.org/10.5506/APhysPolBSupp.10.555}{{\em Acta Phys. Polon.
  Supp.} {\bf 10} (2017)  555}.

\bibitem{Alqahtani:2017jwl}
M.~Alqahtani, M.~Nopoush, R.~Ryblewski, and M.~Strickland, ``{3+1d
  quasiparticle anisotropic hydrodynamics for ultrarelativistic heavy-ion
  collisions},''
\href{http://arxiv.org/abs/1703.05808}{{\tt arXiv:1703.05808 [nucl-th]}}.

\bibitem{Ghiglieri:2018dgf}
J.~Ghiglieri, G.~D. Moore, and D.~Teaney, ``{Second-order Hydrodynamics in QCD
  at NLO},''
\href{http://arxiv.org/abs/1805.02663}{{\tt arXiv:1805.02663 [hep-ph]}}.

\end{thebibliography}\endgroup

\end{document}